\documentclass[fleqn,usenatbib]{mnras}

\usepackage[T1]{fontenc}

\DeclareRobustCommand{\VAN}[3]{#2}
\let\VANthebibliography\thebibliography
\def\thebibliography{\DeclareRobustCommand{\VAN}[3]{##3}\VANthebibliography}

\usepackage{graphicx}	
\usepackage{amsmath}	
\usepackage{amssymb}	

\title[The polarized SED of Centaurus~A]{An X-rays-to-radio investigation of the nuclear polarization from the radio-galaxy Centaurus~A}

\author[F. Marin et al.]{
Fr\'ed\'eric Marin$^{1}$\thanks{E-mail: frederic.marin@astro.unistra.fr},
Thibault Barnouin$^{1}$,
Steven R. Ehlert$^{2}$,
Abel Lawrence Peirson$^{3}$,
\newauthor
Enrique Lopez-Rodriguez$^{4}$,
Maria Petropoulou$^{5}$,
Kinwah Wu$^{6}$,
and Iv\'an Mart\'i-Vidal$^{7,8}$
\\
$^{1}$Universit\'{e} de Strasbourg, CNRS, Observatoire Astronomique de Strasbourg, UMR 7550, 67000 Strasbourg, France\\
$^{2}$NASA Marshall Space Flight Center, Huntsville, AL 35812, USA\\
$^{3}$Department of Physics and Kavli Institute for Particle Astrophysics and Cosmology, Stanford University, Stanford, CA 94305, USA\\
$^{4}$Kavli Institute for Particle Astrophysics \& Cosmology (KIPAC), Stanford University, Stanford, CA 94305, USA\\
$^{5}$Department of Physics, National and Kapodistrian University of Athens, University Campus Zografos, GR 15783, Athens, Greece\\
$^{6}$Mullard Space Science Laboratory, University College London; Holmbury St Mary, Dorking, Surrey RH5 6NT, UK.\\
$^{7}$Departament d’Astronomia i Astrof\'isica, Universitat de Val\`encia, C. Dr. Moliner 50, E-46100 Burjassot, Val\`encia, Spain\\
$^{8}$Observatori Astron\`omic, Universitat de Val\`encia, C. Catedr\'atico Jos\'e Beltr\'an 2, E-46980 Paterna, Val\`encia, Spain
}

\date{Accepted 2023 October 03. Received 2023 September 15; in original form 2023 June 22}

\pubyear{2023}

\begin{document}
\label{firstpage}
\pagerange{\pageref{firstpage}--\pageref{lastpage}}
\maketitle

\begin{abstract}
Centaurus~A is one of the closest radio-galaxies to Earth. Its proximity allowed us to extensively study its active galactic nucleus but the core emission mechanism remains elusive because of local strong dust and gas obscuration. The capability of polarimetry to shave-off contaminating emission has been exploited without success in the near-infrared by previous studies but the very recent measurement of the 2 -- 8~keV polarization by the {\it Imaging X-ray Polarimetry Explorer ({\it IXPE})} brought the question back to the fore. To determine what is the prevalent photon generation mechanism to the multi-wavelength emission from the core of Centaurus~A, 
we retrieved from the archives the panchromatic polarization measurements of the central compact component. We built the total and polarized flux spectral energy distributions of the core and demonstrated that synchrotron self-Compton models nicely fit the polarized flux from the radio to the X-ray band. 
The linear polarization of the synchrotron continuum is perpendicular to the jet radio axis from the optical to the radio band, and parallel to it at higher energies. The observed smooth rotation of the polarization angle in the ultraviolet band is attributed to synchrotron emission from regions that are getting closer to the particle acceleration site, where the orientation of the jet's magnetic fields become perpendicular to the jet axis. This phenomenon support the shock acceleration mechanism for particle acceleration in Centaurus~A, in line with {\it IXPE} observations of several high-synchrotron peak blazars.
\end{abstract}




\section{Introduction}
Centaurus~A (Cen~A) is a nearby \citep[$z$ = 0.00183, $d \approx$ 3.81~Mpc,][]{Harris2010}
radio-loud active galactic nucleus (AGN) that is known for more than 175 years \citep{Herschel1847}. It's optical spectrum classifies it as a type-2 AGN \citep{Dermer1995}, i.e. seen from the edge, where the nuclear light of the central supermassive black hole and its accretion disk is obscured by a thick reservoir of dust and gas known as the ``torus'' \citep{Antonucci1993}. Due to its proximity, Cen~A has been observed at high spatial resolution despite its relatively low bolometric luminosity  ($\sim$ a few 10$^{43}$~erg\;s$^{-1}$, \citealt{Beckmann2011}) with respect to standard type-2 AGNs (a few 10$^{44}$~erg\;s$^{-1}$, \citealt{Lusso2012}). In particular, the complex morphology of the jets and lobes detected from each side of the core has been explored in details. Only apparent in the radio and X-ray bands, those jets become sub-relativistic at a few parsec from the nucleus before expending into plumes at a projected distance of 5~kpc, ending their propagation in space in the form of huge radio lobes extending out to 250~kpc \citep{Israel1998}. The inclination of the inner jets to the observer's line-of-sight is not easy to derive: \citet{Tingay1998}, using the Southern Hemisphere VLBI Experiment (SHEVE) array and the Very Long Baseline Array, estimated that Cen~A's subparsec-scale jet has an inclination of 50$^\circ$ -- 80$^\circ$, while \citet{Muller2014}, using the Southern Hemisphere VLBI monitoring program TANAMI, argued that the inclination is instead likely to be 12$^\circ$ -- 45$^\circ$. This ambiguity in inclination is directly responsible for the various classifications of Cen~A, that is sometimes refereed as a "misdirected" BL~Lac AGN (see \citealt{Bailey1986}) or a radio-galaxy (a radio-loud AGN seen from the side).

The core of Cen~A, however, is not only obscured by the spatially unresolved parsec-scale torus but also by a gigantic warped dust lane which effectively bisects the main body of the host galaxy, partially shrouding the nucleus and all optical structure in the inner 500~pc \citep{Schreier1996}. This dust lane is though to be the remnant signature of a 10$^7$ -- 10$^8$ years old merger of Cen~A's host galaxy with a smaller spiral galaxy \citep{Baade1954,Malin1983,Israel1998} that could be responsible for the activity of the nucleus. The host galaxy of Cen~A is actually a giant, moderately triaxial, elliptical galaxy that contains large amounts of dust, atomic and molecular gas as well as luminous young stars \citep{Israel1998,Espada2009,Espada2012,Espada2017}. The dust lane is observed to be a warped disk, spanning a radius of 1.8 to 6500~pc \citep{Graham1979,Quillen2010,Struve2010}, seen in an edge-on view ($\sim$ 90$^\circ$ $\pm$ 30$^\circ$, \citealt{Quillen1992}). A large-scale magnetic-field of about 3~kpc in diameter tightly follows the warped dust and molecular disk of Cen~A \citep{Lopez-Rodriguez2021}. {\it GALEX} observations revealed an active starburst in the central region obscured by the dust lane ($\sim$ 4``, $\sim$ 60~pc), confirmed by infrared polycyclic aromatic hydrocarbon and silicate features \citep{Alexander1999,Neff2015}. The excess of obscuring material is unfortunate as the nucleus is likely the place where the high energy photons detected by Cherenkov telescopes \citep{Grindlay1975}, {\it EGRET} \citep{Hartman1999}, H.E.S.S. \citep{Aharonian2009} and {\it Fermi} \citep{Abdo2010a} are produced. Determining the mechanisms responsible for the acceleration of charged particles and the radiative processes for producing high-energy photons is fundamental for understanding the physics of cosmic gamma-ray emitters and in particular the role of radio-galaxies as highly efficient relativistic charged particle accelerators. Because of their large number, these would collectively contribute, in a very significant way, to the redistribution of energy in the intergalactic medium \citep{Velzen2012}.

The recent launch of the {\it Imaging X-ray Polarimetry Explorer} ({\it IXPE}, \citealt{Weisskopf2021}) and its observation of Cen~A \citep{Ehlert2022} has shown that high energy polarimetry is able to put constraints on the emission mechanisms responsible for the emission of X and $\gamma$-rays photons. In particular, 
{\it IXPE}'s upper limits on the polarization degree of Cen~A's core allowed to extract kinetic and thermal information about the particle population responsible for generating the X-rays but could not conclude with certainty about the low energy seed photons origin for producing high energy photons. Polarimetry at lower energies, while being scarce in the case of  Cen~A, could not reach a common conclusion and failed at determining the underlying emission mechanism in the core of this radio-loud galaxy \citep{Hough1987,Schreier1996,Packham1996}. The key of such investigation may lie in the panchromatic polarization signatures of the AGN, 
as it was shown in \citet{Marin2018}, \citet{Marin2020}, \citet{Lopez-Rodriguez2018} and \citet{Lopez-Rodriguez2022} that investigating the polarized spectral energy distribution (SED) brings up stronger physical constraints than narrow band polarimetry. 

Motivated by the recent high energy polarization measurement 
by {\it IXPE}, the objective of this paper is to gather all past and recent polarimetric observations of Cen~A in order to try to identify the mechanism (or mechanisms) that governs emission from the core of Cen~A to reveal what is the relativistic electron accelerator process. In Sect.~\ref{SED}, we present our methodology to collect polarimetric data from the core of Cen~A. In Sect.~\ref{Results}, we show the resulting total flux and polarized SEDs, together with the wavelength-dependent polarization degrees and polarization position angles. We discuss our results in the context of past polarimetric investigation in Sect.~\ref{Discussion} before summarizing our findings in Sect.~\ref{Conclusion}.

\section{Building the polarized SED of Cen~A}
\label{SED}
The construction of the polarized SED of Cen~A was based on three distinct axes: 1) collecting data from literature, 2) taking advantage of the brand new and unique X-ray polarimetric measurement by {\it IXPE} and 3) reducing an old yet unpublished ultraviolet polarimetric observation by the Wisconsin Ultraviolet Photo-Polarimeter Experiment (WUPPE, \citealt{Code1993}).

\subsection{Unpublished WUPPE data}
\label{SED:WUPPE}

WUPPE was among the ultraviolet telescopes on the {\it Astro-1} and {\it Astro-2} payloads which flew aboard the space shuttles Columbia and Endeavour, respectively. The first mission lasted nine days (December 2 -- 11, 1990) and the second sixteen days (March 2 -- 18, 1995). It consisted of a 0.5m f/10 classical Cassegrain telescope (area = 1800~cm$^2$, 279~in$^2$) and a spectropolarimeter, where a magnesium flouride Wollaston polarizing beam-splitter was placed between the aperture and the relay mirror and split the beam into two orthogonally-polarized spectra. The polarimetric instrument had a field of view of 3.3' $\times$ 4.4' and a spectral resolution of 6~\AA. Technical details about the instruments are available on the WUPPE operation manual \citep{Nordsieck1994}.

\begin{figure}
\includegraphics[width=8.8cm]{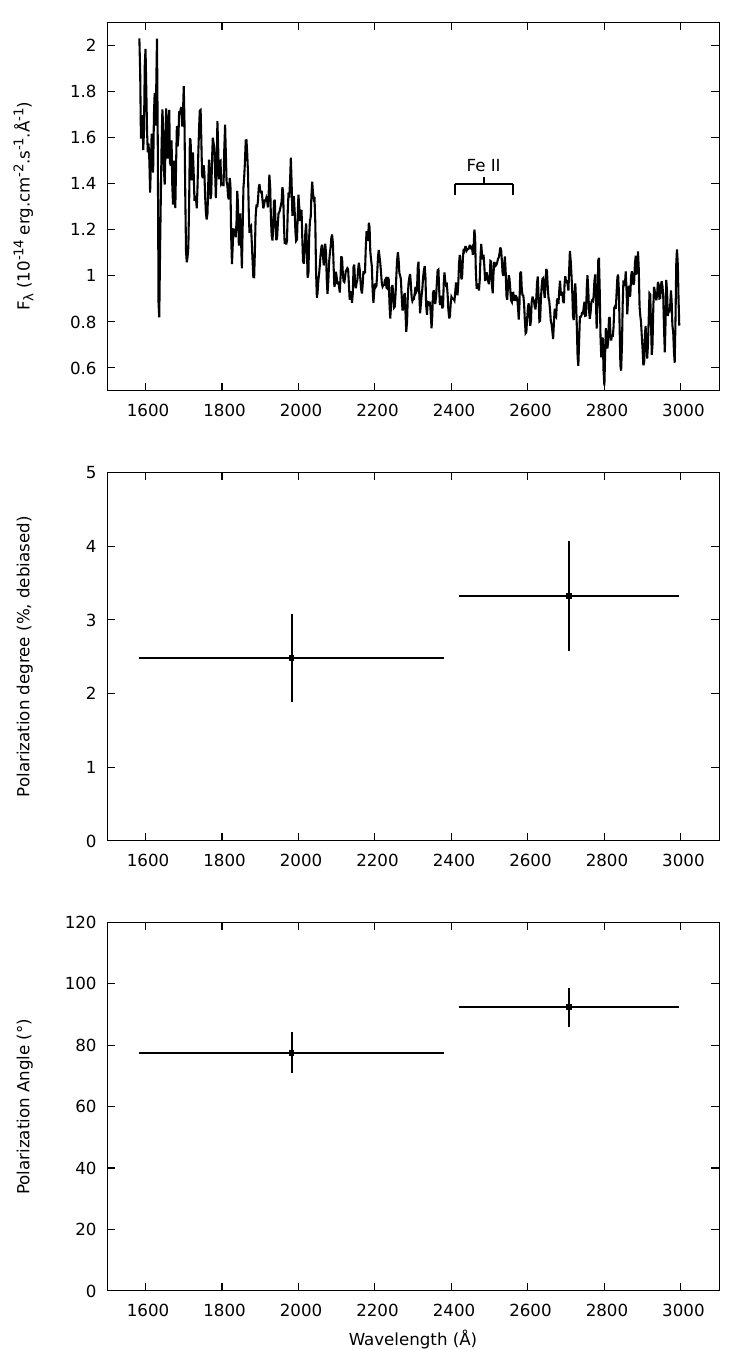}
\caption{WUPPE spectropolarimetry of Cen~A. 
See text for details about observation and data reduction.}
\label{Fig:WUPPE}
\end{figure}

We discovered that WUPPE achieved a single observation of Cen~A that was never published before. We thus summarize the observation and data reduction process hereafter. On March 14, 1995, WUPPE observed Cen~A in spectropolarimetric mode for an useful exposure time (UET) of 1152 seconds. A 6$\arcsec \times$ 12$\arcsec$ aperture was used and the data were acquired between 1450 and 3200~\AA, with a spectral resolution of 16~\AA. Wavelength calibration was achieved using Aluminum lines. Cosmic ray hits, telemetry errors and deviant scans were carefully removed, together with the offsets, thermal background and detector background (\textit{Astro-2 dtr-bkg}). Corrections for thermal drifts were applied and noise was reduced thanks to second order corrections. Flat fields 
for {\it Astro-2} diffuse sources and high gain instrumental polarization calibrations (\textit{wuppe2i}) were applied. Flux calibration was done using 
{\it HST} standards. This leads WUPPE {\it Astro-2} flux calibration to have a relative flux accuracy of 5\% between 1700 and 3000~\AA, and 10\% outside this range. Due to pointing instabilities, fluxes are not absolute. From the instrument manual, the instrumental polarization stability of WUPPE was fixed to $\pm$ 0.04\% and the polarization efficiency was 75\%. WUPPE {\it Astro-2} instrumental polarization is Q =  $-0.04$\% and U =  $-0.08$\%, which has been removed from polarimetric data. Finally, due a detector flaw in the $A$-array, the polarimetric data between about 2380 and 2420~\AA~cannot be trusted and are thus removed from the analysis. We present in Fig.~\ref{Fig:WUPPE} the total flux spectrum (top), polarization degree (middle) and polarization position angle (bottom) as a function of wavelength. The polarization degree has been debiased following the usual procedure $\sqrt{({\rm P}^2-\sigma^2_{\rm P})}$).

The total flux spectrum of Cen~A has a decreasing trend towards longer wavelengths and shows evidence for a relatively smooth feature that is associated with blended Fe~II emission lines \citep{Wills1985}. We checked, using the \textit{dust$\_$extinction} Python package developed by Karl Gordon \citep{Gordon2003}, that this feature does not coincide with the 2175~\AA~ bump seen in the Large Magellanic Cloud, our Milky Way or Milky Way-like galaxies  \citep[see][]{Ma2015}, a prominent feature caused by interstellar extinction. Interestingly, the Fe~II emission feature was absent from the average spectrum of the nuclear region of Cen~A constructed by \citet{Bonatto1996} using all available {\it International Ultraviolet Explorer} ({\it IUE}) spectra of this source (see their Fig.~12). The difference likely arise from the smaller aperture used by WUPPE with respect to the 10$\arcsec$ $\times$ 20$\arcsec$ 
{\it IUE} aperture spectra, better focusing on the AGN core signal. 

The signal-to-noise ratio was unfortunately too low to have a meaningful polarized spectrum. We thus integrated the polarization degree in two energy bands from either side of the 2400~\AA~defect. The debiased polarization degree is 2.48\% $\pm$ 0.60\% (77.6$^\circ$ $\pm$ 6.7$^\circ$) between 1584 and 2380~\AA, and 3.32\% $\pm$ 0.75\% (92.3$^\circ$ $\pm$ 6.3$^\circ$) between 2420 and 2996~\AA. Both values are rather low for a type-2 object, where perpendicular scattering off the AGN polar outflows is expected to produce larger polarization degrees ($>$ 10\%, see e.g., NGC~1068, \citealt{Code1993}). There is not much physical information we can extract from the WUPPE polarimetric observation at this stage. We now need to recontextualize the observation with respect to other wavelength bands (see Sect.~\ref{Results}).

\subsection{Archival data}
\label{SED:Archvies}
We searched the SAO/NASA Astrophysics Data System Abstract Service for all papers mentioning a past polarimetric measurement of the core of Cen~A. We discarded all radio polarization campaigns measuring the total (core plus jets plus lobes) polarization of Cen~A, since the lobes are known to strongly dominate the polarization from the core at most radio frequencies \citep{Burns1983,Goddi2021}. We also obtained from the {\it IXPE} collaboration their very first measurement of the X-ray polarization from Cen~A core. In total, there are only 13 papers reporting a polarization measurement from Cen~A's core, most of them in the near-infrared band. All the papers are listed in Tab.~\ref{Tab:data}. We note a severe lack of measurements in the mid- and far-infrared bands, as well as in the ultraviolet region. In particular, no near-ultraviolet, optical and near-infrared measurements have been achieved since 2000. Nevertheless, the data points cover almost the full electromagnetic spectrum, from the radio to the X-ray band.

\begin{table*}
\centering
\caption{Polarimetric data used to reconstruct the polarized SED of Cen~A, 
	 in descending order of wavelength.}
\label{Tab:data}
\begin{tabular}{lccr} 
	\hline
	Instrument & Wavelength(s) & Aperture & Reference\\
	\hline
	VLA & 1.4~GHz, 4.9~GHz & 3.6$\arcsec$ $\times$ 1.1$\arcsec$ & \citet{Burns1983}\\
	ATCA & 18~GHz & 43$\arcsec$ & \citet{Burke2009}\\
	VLBI (ALMA) & 213 -- 229~GHz & 2.3$\arcsec$ & \citet{Goddi2021}\\
	ALMA & 337 -- 349~GHz & 0.4$\arcsec$ & \citet{Nagai2017}\\
	15-m James Clerk Maxwell Telescope & 800~$\mu$m, 1100~$\mu$m & 15$\arcsec$, 19$\arcsec$ & \citet{Packham1996}\\
	HAWC+ (SOFIA) & 89~$\mu$m & 8$\arcsec$ & \citet{Lopez-Rodriguez2021}\\
	HAWC+ (SOFIA) & 53~$\mu$m & 5$\arcsec$ & \citet{Lopez-Rodriguez2022}\\	
	Hatfield polarimeter (AAT) & J, H, K, L' & 2.25$\arcsec$, 4.5$\arcsec$, 8$\arcsec$ & \citet{Bailey1986}\\
	Hertfordshire polarimeter (AAT) & J, H, K$_n$ & 2.25$\arcsec$, 4.5$\arcsec$, 8$\arcsec$ & \citet{Packham1996}\\
	Hertfordshire polarimeter (AAT) & K & 3.1$\arcsec$ & \citet{Alexander1999}\\
	NICMOS (HST) & 2~$\mu$m & 2.25$\arcsec$ & \citet{Capetti2000}\\
	NSFCAM (IRTF) & 1.65~$\mu$m, 2.2~$\mu$m & 15$\arcsec$, 45$\arcsec$ & \citet{Jones2000}\\
	WFPC (HST) & R' & 0.5$\arcsec$, 1$\arcsec$, 2.25$\arcsec$, 4.5$\arcsec$, 8$\arcsec$ & \citet{Schreier1996}\\
	69-inch Perkins reflector (Lowell observatory) & V & 27$\arcsec$ & \citet{Elvius1964} \\	
	WUPPE & 0.145 -- 0.32~$\mu$m & 6$\arcsec$ & this paper \\
	IXPE & 2 -- 8~keV & 45$\arcsec$ & \citet{Ehlert2022}\\
	\hline
\end{tabular}
\end{table*}

\section{results}
\label{Results}

\subsection{The total flux and polarized SEDs}
\label{Results:SED}
\begin{figure}
\includegraphics[width=8.8cm]{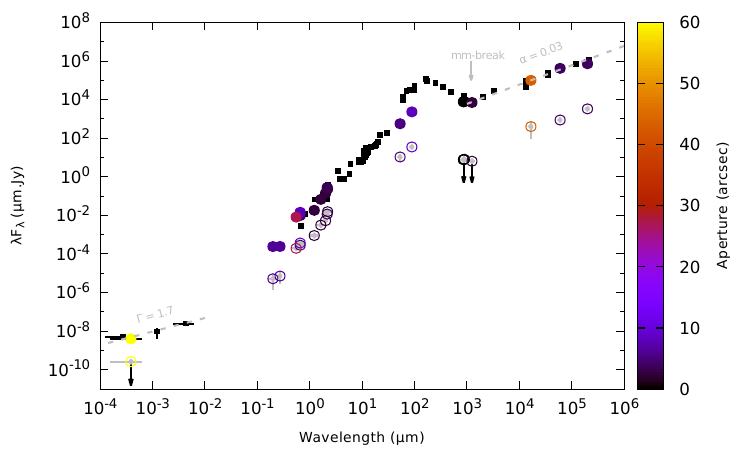}
\caption{X-ray to radio measurements of the total and polarized fluxes of Cen~A. 
	The instrument apertures used to measure the fluxes are color-coded.
	Black-filled squared symbols indicate total flux data taken from NED, while 
	circular symbols indicate flux measurements from polarimetric observations.
	Polarized fluxes are differentiated from total fluxes using a central gray-filled 
	circle surrounded by an empty colored-circle.}
\label{Fig:TF_PF}
\end{figure}

We present the total and polarized flux SEDs of Cen~A in Fig.~\ref{Fig:TF_PF}. In addition to the total fluxes collected in the polarimetric papers, we added core flux measurements from the NASA/IPAC Extragalactic Database. We only kept flux measurements within the first 50$\arcsec$ of Cen~A in order to remove as much lobe and jet contamination as possible. When several data points were available for the same wavelength, we kept the one with the smallest aperture. 

The total flux SED (in $\lambda$F$_\lambda$) has been presented and studied in numerous papers \citep[e.g.][]{Alexander1999,Rieger2009,Petropoulou2014}, we therefore content ourselves with describing it succinctly to dwell at greater length on its polarized properties. The total flux SED appears to monotonically decrease from the radio band to the millimeter band, a common signature for synchrotron emission in radio-loud AGNs \citep{Padovani2016}. The non-thermal synchrotron power-law spectral index $\alpha$ measured for the core radio emission ($\alpha \approx$ 0.03 on Fig.~\ref{Fig:TF_PF}) differs from the one measured in the jets and lobes ($\alpha \approx -0.7$, \citealt{Alvarez2000}). The latter value is consistent with optically thin synchrotron emission from freshly accelerated electrons in the large-scale components, and the flatter spectral slope of the core emission is likely caused by synchrotron self-absorption (i.e. optically thick) in the core and at the base of the jets \citep[see][]{Eckart1986,Blandford1979}. Around 1200~$\mu$m (1.2~mm), the total flux spectrum is characterized by a break which marks the transition between synchrotron and dust plus starburst emission. Indeed, Cen~A is known to harbor active star formation sites within its dust lane and in regions beyond the bulk of the molecular material, resulting in strong, black-body-like emission in the 100 -- 1200~$\mu$m waveband \citep{Unger2000}. At wavelengths smaller than 100~$\mu$m, the starburst contribution decreases and infrared re-emission from the AGN torus prevails down to about 1~$\mu$m \citep{Alexander1999}. At optical and near-ultraviolet wavelengths, we do not detect the usual Big Blue Bump, a characteristic of AGNs. This was already reported by \citet{Rieger2009}, and it indicates that either the standard accretion disk scenario is not applicable in the case of Cen~A (see, e.g., \citealt{Marconi2001}) or that optically thick thermal emission is not the dominant mechanism to explain the observed optical and near-ultraviolet continua. Finally, the X-ray continuum of Cen~A can be fitted by a single power-law spectrum with a photon index $\Gamma \approx 1.7-1.9$ from 3 to 120~keV, the 0.1 -- 2~keV spectrum of the nucleus being heavily absorbed \citep{Hall1976,Mushotzky1978,Turner1997,Furst2016,Ehlert2022}. 

The polarized SED is shown using empty circles on Fig.~\ref{Fig:TF_PF}. Unfortunately, due to the scarcity of polarimetric measurements, only a handful of data point are available. However, they do cover almost all the electromagnetic spectrum, which allow us to examine its properties. The polarized SED follows the shape of the total flux SED, but the difference between the two is much larger in the radio and millimeter bands than in the near-infrared and optical bands. On one hand, this increase of polarization with frequency is something that is typically observed in blazars \citep{Bjornsson1985,Jorstad2006,Liodakis2022} and could be intrinsic to the jet itself. On the other hand, several other regions could produce the observed polarization by emission, absorption and/or scattering, while contributing to the observed total flux too. There are only upper limits for the polarized flux around the mm-break, which is related to the non-detection of polarization at levels $\ge$ 0.07\% \citet{Nagai2017}. The upper limit placed by IXPE on the polarized flux of Cen~A indicates that no dramatic increase of the polarized flux is expected in the high energy band.

\subsection{Polarization degree and angle}
\label{Results:Pol}
\begin{figure}
\includegraphics[width=8.8cm]{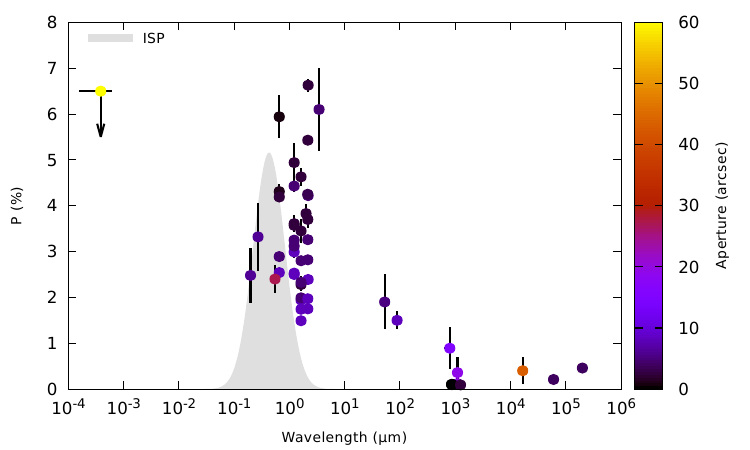}
\caption{Same as Fig.~\ref{Fig:TF_PF}, but for the linear polarization degree P of the continuum.
	The transparent gray curve indicates the level of contamination from Cen~A dust lane and 
	interstellar polarization (ISP) as measured by \citet{Hough1987}. Note : IXPE measurement
	was achieved using a 1' diameter aperture}
\label{Fig:PO}
\end{figure}

\begin{figure}
\includegraphics[width=8.8cm]{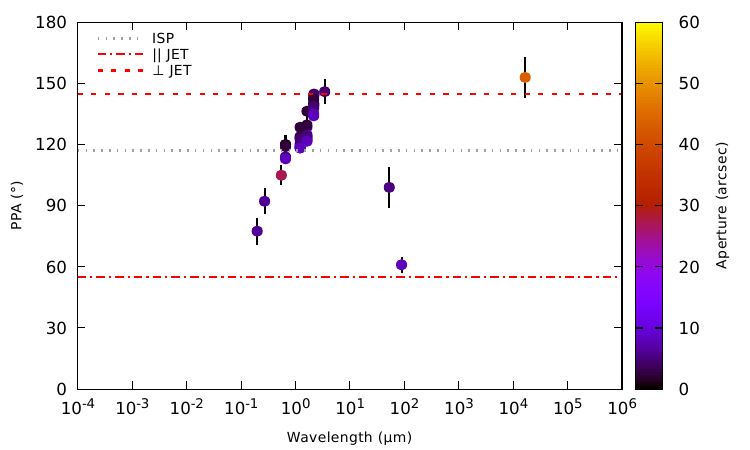}
\caption{Same as Fig.~\ref{Fig:TF_PF}, but for the polarization position angle (PPA) of the continuum.
	The gray dashed line indicates the level of contamination from Cen~A dust lane and 
	interstellar polarization (ISP), associated to a polarization angle of 117$^\circ$ \citep{Hough1987}. 
	The red dashed lines correspond to the orientation of the polarization position angle 
	(either parallel or perpendicular) with respect to the position angle of the AGN radio jet, 
	measured to be 55$^\circ$ $\pm$ 7$^\circ$ \citep{Burns1983}.}
\label{Fig:PA}
\end{figure}

To better break down the polarized SED of Cen~A, we present the broadband polarization measurements in Figs.~\ref{Fig:PO} and \ref{Fig:PA} for the continuum linear polarization degree (P) and polarization position angle (PPA), respectively. On both figures, we added in gray the contribution of polarized light by the dust lane in Cen~A and the interstellar medium towards the AGN (both grouped under the acronym ISP for brevity). Indeed, \citet{Hough1987} measured the optical and infrared linear polarization produced by transmission of radiation through the dust lane of Cen~A's galaxy and found that they were consistent with the standard form of interstellar polarization used in our Galaxy with values P$_{\rm max}$ = 5.16\% $\pm$ 0.04\% (PPA = 117$^\circ$) at $\lambda_{\rm max}$ = 0.43~$\mu$m. The foreground material in our Galaxy was found to marginally contribute to the observed polarization by less than 0.5\%. On Fig.~\ref{Fig:PA}, the position angle of the radio jet (PA, measured to be 55$^\circ$ $\pm$ 7$^\circ$, \citealt{Burns1983}) is also marked using red dashed lines. It is common to compare the observed PPA to the radio PA of the jet to determine the geometrical distribution of scatterers in radio-quiet AGNs. By doing so, \citet{Antonucci1983} and many other authors have proven that the vast majority of Seyfert-1s have an optical polarization angle parallel to the jet PA, while Seyfert-2s have their optical PPA perpendicular to the jet PA, providing crucial information on the matter distribution in the innermost regions of non-jetted AGNs.

The polarization degree appears to be strongly wavelength-dependent across the wavelengths from radio to X-rays. In the radio band, the polarization degree is low (with P $<$ 1\%) and the associated PPA could not be measured due to strong Faraday rotation toward the core of Cen~A. The only exception is at 18~GHz, where the observed polarization angle is perpendicular to the jet direction, although the rather large aperture (43$\arcsec$) used to integrate the polarized flux implies that jet emission partly contribute to the observed PPA. Such low polarization degrees in the radio band can be attributed to two possible mechanisms: 1) depolarization by Faraday rotation in radiatively inefficient accretion flows around the central supermassive black hole \citep{Yuan2003} or 2) synchrotron emission in 
the optically thick regime \citep[see e.g.][]{Ginzburg1969}, perhaps in addition, with some dilution caused by disorganised magnetic fields within the emission region. The first hypothesis is wavelength-dependent and thus cannot explain the almost uniform, P $<$ 1\%, core radio polarization. However, the second scenario is consistent with the spectral slope of the total flux SED. 

At the position of the mm-break, the core polarization of Cen~A becomes consistent with zero and then rises again as the far-infrared dust and starburst emission onsets. This is a typical signature of a transition region, where a new process starts to dominate over a former one. Here, we attribute the mm-break and the vanishing of the polarization degree to the transition from synchrotron at radio wavelengths to thermal polarized emission in the far-infrared. In fact, the situation is very similar to the case of Cyg~A that was studied by \citet{Lopez-Rodriguez2018}. The radio band of Cyg~A is dominated by synchrotron emission with a flat core spectrum and a turnover is observed in the (sub)millimeter band. This is coherent with the fact that synchrotron polarization in Cen~A is very low and thus easily masked by another source of polarized light. \citet{Lopez-Rodriguez2022} showed that the far-infrared (89~$\mu$m) polarized emission is predominantly arising from polarized thermal emission by means of magnetically aligned dust grains. These authors also provided an upper limit on the synchrotron polarization of $<$ 0.4\% at millimeter wavelengths using a total and polarized SED fitting.
 
The fact that below 1000~$\mu$m the polarization degree starts to rise from 1\% to 6\% in the near-infrared indicates that another polarization mechanism, different from synchrotron emission, onsets. The infrared band wavelength-dependence and the rise of P with decreasing wavelengths is the tell-tale signature of polarization due to dichroic absorption and emission by magnetically aligned grains in dusty discs \citep{Efstathiou1997,Dotson2000}. In fact, this wavelength dependence of polarization can be directly related to the properties of the dusty region: the grains shape, the optical depth and the degree and orientation of the particles. Two notable signatures of dichroic absorption and emission are a bump in polarization around 10~$\mu$m and a rotation of the polarization position angle at a wavelength that depends on the opacity of the medium. If there are no measurements made around 10~$\mu$m allowing us to verify the first statement, we can see a rotation of the angle of polarization between the measurements taken at 53~$\mu$m and 89~$\mu$m. This seems to support the dichroic polarization hypothesis, already strongly supported by the polarization behavior in the infrared band, but new polarization measurements between 2~$\mu$m and 50~$\mu$m are needed to check the presence of the polarization peak around of 10~$\mu$m and better characterize the variation of the PPA. The current far-infrared data, taken from \citet{Lopez-Rodriguez2021} and \citet{Lopez-Rodriguez2022}, are diluted by the emission and/or absorption from the dust lane in the host galaxy, so the recorded polarization degrees are lower than the intrinsic polarization from the nucleus. The authors estimated that the intrinsic far-infrared polarization of Cen~A's core reaches 4 -- 5\%, while the polarization angles remain unchanged, further strengthening the hypothesis that, in the mid-to-far infrared band, polarization mainly rises from magnetically aligned dust grains.

Below 2~$\mu$m, polarization resulting from dichroism is less likely to happen due to its inefficiency to produce the expected levels of P \citep{Hough1987,Packham1996}. Scattering is an obvious candidate, although synchrotron emission could play a role too. The fact that the PPA is perpendicular to the radio jet at 2~$\mu$m (as expected from Seyfert-2 galaxies) tends to indicate that polar scattering is the principal contributor to the observed polarization. If synchrotron emission alone could produce the $\sim$ 6\% polarization observed in the optical band, it also should be detectable in the millimeter, which is not the case, although the contribution of starburst light could be the reason for the upper limits measured around the mm-break \citep{Packham1996}. Because of the wavelength-dependence of P in the optical and near-ultraviolet bands, Mie scattering on polar dust is more probable than electron scattering, the latter being wavelength-independent (see \citealt{Antonucci1994} and \citealt{Marin2018} for examples about NGC~1068). However, we note the most intriguing behavior for the polarization angle. It appears to gradually rotate from an angle perpendicular to the jet at 2~$\mu$m to an angle that is parallel at 0.2~$\mu$m. If we interpolate the data, the PPA would reach a value of 55$^\circ$ around 0.06~$\mu$m $\pm$ 0.02~$\mu$m ($\sim$ 20~eV). What is the physical process behind such transition? This question will be answered in the next subsection.

Finally, the polarimetric measurement achieved at high energies, provided by {\it IXPE} in the 2 -- 8~keV band, is only an upper limit with no associated PPA. Although it is consistent with Compton scattering levels of polarization \citep{Bonometto1973,Poutanen1994,Celotti1994,Peirson2019}, the specific seed photon population responsible for production of the X-rays could not be identified from {\it IXPE} alone \citep{Ehlert2022}.

\subsection{Fitting the total flux and polarized SEDs}
\label{Results:Fit}
\begin{figure}
\includegraphics[width=8.8cm]{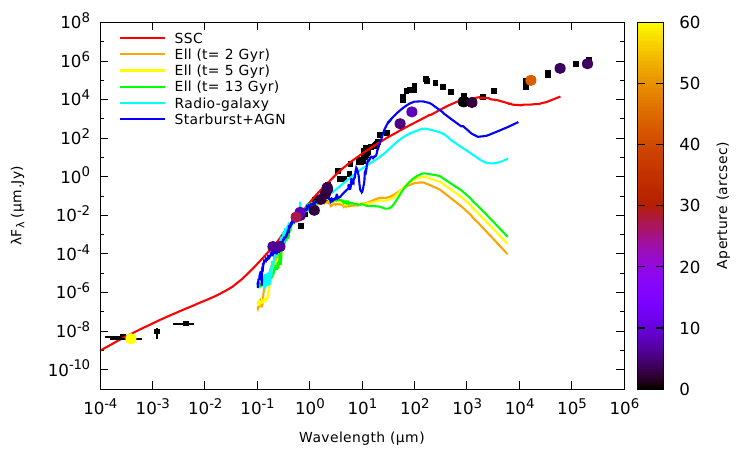}
\caption{Total flux SED of Cen~A (see Fig.~\ref{Fig:TF_PF} for details), together with
	 several SEDs from literature. The SSC model is from \citet{Abdo2010} and \citet{Petropoulou2014}, 
	 while the other spectra are extracted from the SWIRE Template Library \citep{Lonsdale2004,Polletta2007}.
	 See text for details.}
\label{Fig:Fit_TF}
\end{figure}

The panchromatic polarization signatures of Cen~A clearly indicate that several mechanisms are responsible for the observed polarization levels. In contrary to jet-dominated AGNs, such as blazars, synchrotron+Compton emission does not prevails from the radio to the X/$\gamma$-rays in radio-galaxies \citep{Potter2012,Potter2013a,Potter2013b,Potter2013c}. To test this, we tried to fit the total flux SED of Cen~A with various templates from the literature. First, we used the most up-to-date SED modeling of Cen~A that was developed by \citet{Abdo2010}, later modified by \citet{Petropoulou2014}, and based on the synchrotron self Compton mechanism (SSC). In this modified one-zone SSC model, relativistic electrons are responsible for the radiation observed up to the GeV energy range, while the GeV-TeV emission itself is attributed to photohadronic processes to correctly fit the gamma-ray band. We also used the SWIRE Template Library \citep{Lonsdale2004,Polletta2007} as it contains numerous SEDs of elliptical galaxies (such as the host of Cen~A), starbursts galaxies, type-1 and type-2 radio-loud and radio-quiet AGNs, as well as composite SEDs (starburst+AGN) covering the wavelength range between 0.1~$\mu$m and 1000~$\mu$m. Our goal is not to make a spectral decomposition but rather identify the weakness of each model. The underlying reason will become clear in the next paragraph. The results are shown in Fig.~\ref{Fig:Fit_TF}. All template SEDs have been normalized at 0.55~$\mu$m (as in the SWIRE Template Library). It is clear that none of the observed nor modeled SEDs get close to the total flux SED of Cen~A. Elliptical galaxies only match the observed fluxes in the optical band, the radio-galaxy template better fits the data up to the infrared but starts to fail predicting the millimeter and radio bands data, and the starburst+AGN template only gets closer to the millimeter emission. The SSC model developed by \citet{Abdo2010} and later modified by \citet{Petropoulou2014}, that was meant to fit the high-energy (gamma-ray) SED of Cen~A, does not match the observed millimeter and radio bands data, probably due to the lack of inhomogeneous modelling (see, e.g., \citealt{Ghisellini1985}) and to the fact that synchrotron emission is likely to be partially extinguished at those wavelengths. The SSC model neither provide a very good fit to very soft X-rays, where line-of-sight absorption was not accounted for. All these points were  detectable in Fig.~1 of \citet{Petropoulou2014} and are evident in our Fig.~\ref{Fig:Fit_TF}. It is thus clear that Cen~A core's total flux SED is a composite of many emission mechanisms/regions, supporting our conclusion from Sect.~\ref{Results:Pol}.

However, one question remains unanswered: if we could strip-off Cen~A from its dust cocoon, thereby eliminating the dust lane and the starburst regions, what would be the true physical process(es) governing its core emission? This question is of fundamental importance since it could help us to solve the origin of the apparent stochastic variation of AGN cores, constrain the acceleration and radiative mechanisms that boost particles to ultra-high energies, and better probe the innermost regions that surround supermassive black holes. It is physically impossible to achieve so using the total flux SED since it is polluted by various emitting regions not related to the core continuum. The polarized flux SED, on the contrary, shaves-off unpolarized emission. However, we note that, at optical and infrared wavelengths, the dust lane still extinguishes the intrinsic polarization of the core, so the polarized flux needs to be scaled to the contribution of the starlight and emission from the dust lane along the line-of-sight to reveal its true polarization degree. For this reason, we will concentrate of the shape of the whole polarized SED with respect to the total flux SED.

\begin{figure}
\includegraphics[width=8.8cm]{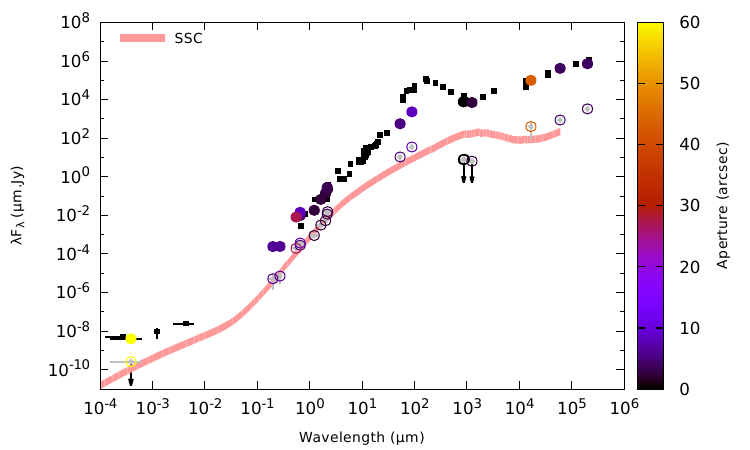}
\caption{Same as Fig.~\ref{Fig:Fit_TF} but for the polarized SED.
	Only the SSC model is shown using a semi-transparent red
	filled curve, with upper and lower values set to 2.4\% and 
	1\% of the total flux SSC model, respectively.}
\label{Fig:Fit_PF}
\end{figure}

The polarized SED of Cen~A's core together with the SSC model can be seen in Fig.~\ref{Fig:Fit_PF}. Minimizing the sum of squared residuals gives a normalization factor of 0.024 for the SSC (that was meant to model the total flux SED) to match the polarized SED. But because polarization is a positive quantity, the normalization factor goes down to 0.01 if we want to be sure that every single data point (excluding upper limits) is situated on or above the theoretical curve. In both cases, the model appears to fit with a great precision the polarized SED from the radio to the X-ray band, much better than what was postulated for the total flux. Even more convincing, the wavelength at which the inverse Compton total flux peak of the SSC model starts to dominate over the total flux peak of synchrotron emission happens around 0.04~$\mu$m. This is in very good agreement with the extrapolated value for the end of the PPA rotation (see Fig.~\ref{Fig:PA}). The only apparent ''downside`` of the fit are the millimeter polarized data, but it is easily explained: because there is a competition between synchrotron emission (with polarization perpendicular to the jet) and emission by magnetically aligned dust grains (with parallel polarization, see data at 89~$\mu$m on Fig.~\ref{Fig:PA}), the resultant of the two vector values is null and therefore the polarized flux tends to zero.

Based on the {\it IXPE} observations of Mrk~421 \citep{Gesu2022} and Mrk~501 \citep{Liodakis2022}, we have come to a consensus that shocks are the most likely mechanism for accelerate charged particles (which in turn produce X-rays) in high-synchrotron peak blazars. A tell-tail signature is that the shock would compress and align the magnetic fields crossing the shock, resulting in an ordered configuration of fields perpendicular to the shock normal \citep{Marscher2014,Peirson2019}. The observed polarization angle should then be aligned with the jet, such as it appears to be the case for the extrapolated far-ultraviolet emission in Cen~A. It implies that the ultraviolet photons of Cen~A are most likely high energy synchrotron photons emitted close to the acceleration region.

The normalization factor of the polarized SED is directly related to the polarization degree associated with the synchrotron component. Hence, without the addition of a diluting component (such as in the mm-break), we postulate that the expected level of synchrotron polarization from the core of Cen~A is of the order of 1--3\%. The synchrotron continuum polarization likely varies with frequencies, as in blazars, which explains why the SSC fit is not perfect; but here it is the adequacy of the polarized SED and the panchromatic model which prevails. Any observed polarization degree higher than a few percents most likely results from the addition of another polarized component, such as dichroic absorption and emission by magnetically aligned dust grains in the circumnuclear disk obscuring the AGN in the infrared band ($\sim$50 -- 500~$\mu$m, \citealt{Lopez-Rodriguez2022}). Furthermore, in the SSC model, the polarization degree of the Compton up-scattered X-rays is related to the synchrotron radiation acting as seed photons for Compton scattering \citep{Peirson2019}. It is predicted to be 2 -- 5 times smaller than its low-energy counterpart, placing a limit on the expected X-ray polarization of Cen~A below 1\%. This explain the non-detection by {\it IXPE}, which had a minimum detectable polarization at 99\% confidence level of 6.5\%).

In conclusion, stripping off the core of Cen~A thanks to polarized flux reveals that the SSC model is correct. It follows that the seed photon population for all the reprocessed emission (dust reemission in the mid and far-infrared, scattering in the near-infrared, dichroic transmission in the optical, inverse Compton scattering in the X-rays) likely originates from synchrotron emission from the AGN core.

\section{Comparison to previous narrow-band polarimetric observations}
\label{Discussion}
Cen~A polarization has been measured since the 80's \citep{Hough1987}, with great focus on the near-infrared band (1 -- 2~$\mu$m) predominantly. The optical band was avoided as much as possible due to heavy contamination by the dust lane and insufficient spatial resolution \citep{Elvius1964}. The far-infrared\footnote{But this has changed in the past several years thanks to HAWC+/SOFIA which shows impressive results in the field of AGNs. In particular, radio-loud AGNs are found to be highly polarized from magnetically aligned dust grains in the dusty obscuring material surrounding the AGN at far-infrared wavelengths \citep{Lopez-Rodriguez2021,Lopez-Rodriguez2022}.}, far-ultraviolet and X-ray bands did not have dedicated polarimeters for decades and only a couple of measurements have been achieved. Millimeter polarization measurements are consistent with zero and radio polarimetry is completely dominated by the jets and lobes polarization. Most, if not all, current conclusions about the physical processes responsible for the detected polarization in Cen~A come from near-infrared polarimetry. We now revisit all the near-infrared papers on the matter and replace their conclusions in the context of the panchromatic polarized SED that was shown in Figs.~\ref{Fig:TF_PF} and \ref{Fig:Fit_TF}.

\citet{Hough1987} were the first to measure the near-infrared polarization of Cen~A using multi-apertures photo-polarimetry. Their large aperture measurements were completely polluted by the dust lane (as evident from the polarization position angle parallel to the warped dusty structure), but smaller aperture polarimetry showed higher polarization degrees and a swing of the PPA towards a perpendicular orientation with respect to the radio jet. They corrected their polarization measurement for the diluting effect of starlight and found that the intrinsic polarization was in the range 8.8 -- 11.1\% for the K band and 7.2 -- 15.4\% for the H band. They interpreted  
the slow rise of polarization towards the blue band as a counter-argument for dichroic absorption, a conclusion we share. To satisfy the energy balance requirement and the shape of the SED, non-thermal radiation is assumed over thermal radiation \citep{Grasdalen1976} and \citet{Hough1987} concluded that synchrotron emission in the jet was responsible for the high polarization of Cen~A's core. We have shown that non-thermal radiation is indeed favored at the core of Cen~A, but the synchrotron polarization is likely to have an intrinsic value of $\sim$ 1\%. However, core synchrotron photons are the most obvious candidates to act as seeds for polar scattering, increasing the observed polarization following the Malus's law. 

Using the {\it HST}/WFPC, \citet{Schreier1996} obtained polarimetric images of Cen~A in the R' and I' bands at a 0.2$\arcsec$ resolution. They discovered that the peak of infrared polarization is coincident with a compact emission knot very close to the nucleus position. This knot also appears to be slightly less obscured than the surrounding regions in the optical band. Looking at this knot in particular, \citet{Schreier1996} found a relatively similar polarization degree and polarization angle to the infrared measurements from \citet{Hough1987}. This dismisses dichroism as the main polarizing mechanism in the near-infrared band. Because neither morphological nor emitting features associated to the radio jets are detected in optical images and polarization maps, the most likely scenario for the production of the high polarization degrees from the knot is scattering onto polar material, which we also support. 

Almost at the same time, \citet{Packham1996} presented J, H and K band polarization maps using the Anglo-Australian Telescope's near-infrared camera IRAS, as well as millimeter polarimetric observations. The nuclear polarization in the K band is clearly different from the ones obtained at shorter wavelengths, strengthening the conclusion that dichroic emission from the dust lane pollutes the AGN intrinsic polarization in the optical band. The additional polarization component at 2.2~$\mu$m, with a PPA distinctively different from ISP's polarization angle, was not attributed to synchrotron emission because of the null polarization in the millimeter band. The reasoning was however partly incorrect because the authors compared the flux and polarization of Cen~A's core to blazars such as 3C~279 or OJ~287. Nevertheless, modeling of their results strongly tip the scale towards scattering and \citet{Packham1996} concluded that the observed near-infrared flux from the central source in Cen~A arises entirely from scattering onto small dust grains or electrons. Interestingly, they mention that the radiation scattered towards us could find its origin from synchrotron emission.

\citet{Alexander1999} achieved K-band spectropolarimetric observations of the nuclear region of Cen~A and failed to detect broad Br$\gamma$ in either total or polarized flux. The synchrotron hypothesis was not investigated (even though it would produce a featureless polarized continuum) and the authors only investigated the dichroic and scattering scenarios. In both cases, the expected equivalent width and flux of the broad Br$\gamma$ line is hypothesized to be very weak and the authors could not conclude on the most favorable scenario. In the light of the synchrotron seed for scattering scenario, it appears logical that the total and polarized flux spectra are featureless.  

Using the 1 -- 5~$\mu$m imager with a 256 $\times$ 256 InSb detector NSFCam on the NASA {\it Infrared Telescope Facility} ({\it IRTF}), \citet{Jones2000} achieved imaging polarimetry at 1.65 and 2.2~$\mu$m. He produced large scale (45'') and small scale (15'') squared aperture polarization maps of Cen~A to probe its magnetic field geometry. The polarization position angle was found to be mainly aligned along the dust lane, with a rotation of the angle at the position of the nucleus. This twist was attributed to the combined effect of the intrinsic polarization for the nucleus and the interstellar polarization in the intervening dust lane being at different angles. The nice polarization maps of \citet{Jones2000} could not be used in this work since the author did not provide integrated polarization degree for the core of Cen~A. No Stokes Q and U maps were provided either, preventing us to estimate the polarization degree of the central region. 

Finally, \citet{Capetti2000} performed the last historical near-infrared polarimetric observation of Cen~A at the beginning of the millennium using the multi-object spectrometer NICMOS on-board {\it HST}. They found that the large scale polarization component is consistent with dichroic transmission through the dust lane, but the core polarization is most likely originating from scattering onto optically thin matter. In particular, the polarized flux 
continuity from radio to near-infrared put the scattering scenario on a bi-conical medium close to the nucleus more likely than dichroic transmission as the latter would require a 50~mag extinction to explain the observed polarization degree. A synchrotron origin for the infrared polarization is discussed but dismissed with respect to scattering.

\section{Conclusions}
\label{Conclusion}

We have shown that a simple SSC model, in which electrons are accelerated to relativistic energies through shocks, could reproduce the polarized SED provided its emission is only weakly polarized, i.e. that the magnetic fields are significantly disordered on the emission scales. These accelerated electrons interact with the entangled magnetic field and emit synchrotron radiation in the radio-to-ultraviolet bands, acting as the seed photons for inverse Compton processes that are responsible for the observed X-rays. Among the signatures we have identified are the wavelength-dependent polarization degree and the rotation of the polarization position angle in the ultraviolet, from perpendicular to parallel to the jet axis. Our results are consistent with the measured polarization degree and angle in blazars at X-ray energies by {\it IXPE}, allowing us to estimate that the X-ray polarization of Cen~A is likely of the order of, or below, 1\%.

The construction of the polarized SED of Cen~A also allowed us to identify the contribution of the various emitting, absorbing or scattering regions that are situated around the AGN core :
\begin{itemize}
  \item \textbf{Radio}: synchrotron emission from the core dominates the SED, with a low ($<$ 1\%) polarization degree due to the optically thick nature of the emission region and the associated polarization position angle may be perpendicular to the jet axis. Once jets and lobes are accounted for, they completely overwhelm the core signatures. 
  \item \textbf{Millimeter}: transition region from synchrotron at radio wavelengths to thermal polarized emission in the far-infrared. The polarization degree is essentially null. 
  \item \textbf{Far-infrared}: the total and polarized SED follows the thermal emission bump likely arising from the dust surrounding the AGN. The inferred magnetic field is found to be perpendicular to the radio jet and the polarization degree points towards magnetically aligned dust grains at 53 -- 89~$\mu$m.  
  \item \textbf{Mid-infrared}: dichroic absorption modifies the core synchrotron polarization and imposes a polarization angle parallel to the jet axis
  \item \textbf{Near-infrared and optical}: scattering of synchrotron seed core photons off the AGN polar outflows produces high polarization degrees associated with a polarization position angle perpendicular to the jet axis. Where dust dominates over the AGN outflows, a large fraction of the observed polarization is due photons traveling through the dust lanes, polarized by dichroic transmission from aligned dust grains. 
  \item \textbf{Ultraviolet}: The polarization position angle rotates and aligns with the jet axis. The polarized emission most likely comes from synchrotron emission from regions close to the acceleration shocks, where the magnetic fields become perpendicular to the jet.
  \item \textbf{X-rays}: inverse-Compton scattering dominates. The polarization angle is predicted to be parallel to the jet axis but the polarization degree is yet to be measured. We constrain its polarization degree to, or below, 1\%.
\end{itemize}

One last question remains unanswered about the central engine of Cen~A : is it powered by radiatively efficient black hole accretion? At optical and near-ultraviolet wavelengths, the famous Big Blue Bump that is characteristic of AGNs is not detected. Determining if such an accretion structure exists is fundamental to understand the launching mechanisms of jet and the production of high energy photons, and it is directly related with one of the long-lasting opened question in the field of AGNs: does Cen~A, and by extension other FR-I radio galaxies, really host a type-1 AGN hidden by optically thick nuclear dust, as predicted by AGN unification? We have shown that the nuclear optical/NIR emission is dominated by synchrotron emission from the base of the radio jet, but is there little accretion or not accretion at all? High-quality spectropolarimetric observations are needed. Measuring the polarization of the Balmer emission line is crucial for determining if the core emission is only synchrotron or if there is also a thermal component. In case of a thermal origin, the H$\alpha$ line will be significantly broadened in polarized flux and have a polarization position angle perpendicular to the jet axis. On the contrary, a synchrotron origin will not broaden the line's full width half maximum in polarized flux and its polarization angle could have any orientation. Spectropolarimetry can thus reveal the presence of radiatively efficient black hole accretion despite obscuration thanks to perpendicular scattering onto polar material.

\section*{Acknowledgements}
The authors thank the anonymous referees for his/her helpful comments that improved the quality of the manuscript. The {\it Imaging X ray Polarimetry Explorer} ({\it IXPE}) is a joint US and Italian mission.  The US contribution is supported by the National Aeronautics and Space Administration (NASA) and led and managed by its Marshall Space Flight Center (MSFC), with industry partner Ball Aerospace (contract NNM15AA18C).  The Italian contribution is supported by the Italian Space Agency (Agenzia Spaziale Italiana, ASI) through contract ASI-OHBI-2017-12-I.0, agreements ASI-INAF-2017-12-H0 and ASI-INFN-2017.13-H0, and its Space Science Data Center (SSDC) with agreements ASI-INAF-2022-14-HH.0 and ASI-INFN 2021-43-HH.0, and by the Istituto Nazionale di Astrofisica (INAF) and the Istituto Nazionale di Fisica Nucleare (INFN) in Italy.  This research used data products provided by the IXPE Team (MSFC, SSDC, INAF, and INFN) and distributed with additional software tools by the High-Energy Astrophysics Science Archive Research Center (HEASARC), at NASA Goddard Space Flight Center (GSFC).

\section*{Data availability}
The data underlying this article are freely available via the SAO/NASA Astrophysics Data System Abstract Service, at https://ui.adsabs.harvard.edu/. The datasets were derived from sources in the public domain listed in Tab.~\ref{Tab:data}.

\bibliographystyle{mnras}
\bibliography{bibliography} 

\bsp	
\label{lastpage}
\end{document}